\documentclass[twocolumn]{aastex631}

\usepackage{{xcolor}}
\definecolor{grey}{HTML}{ababab}
\definecolor{grey_light}{HTML}{d1d1d1}

\newcommand{\cai}{Ca \textsc{i}}
\newcommand{\caii}{Ca \textsc{ii}}
\newcommand{\mgii}{Mg \textsc{ii}}
\newcommand{\caiii}{Ca \textsc{iii}}

\newcommand{\hazel}{\texttt{Hazel2}}

\usepackage{cancel}
\usepackage{color}

\definecolor{deepmagenta}{rgb}{0.8, 0.0, 0.8}
\definecolor{green}{rgb}{0.0, 0.5, 0.1}

\shorttitle{Accelerating non-LTE synthesis and inversions
with graph networks}
\shortauthors{Vicente Ar\'evalo et al.}
\graphicspath{{./}{figures/}}

\begin{document}

\title{Accelerating non-LTE synthesis and inversions
with graph networks}

\author[ 0000-0003-3896-836X ]{A. Vicente Ar\'evalo}
\affiliation{Instituto de Astrofísica de Canarias, C/Vía Láctea s/n, E-38205 La Laguna, Tenerife, Spain}
\affiliation{Departamento de Astrofísica, Universidad de La Laguna, E-38206 La Laguna, Tenerife, Spain}

\author{A. Asensio Ramos}
\affiliation{Instituto de Astrofísica de Canarias, C/Vía Láctea s/n, E-38205 La Laguna, Tenerife, Spain}

\author{S. Esteban Pozuelo}
\affiliation{Instituto de Astrofísica de Canarias, C/Vía Láctea s/n, E-38205 La Laguna, Tenerife, Spain}



\begin{abstract}

The computational cost of fast non-LTE synthesis is one of the challenges that limits the development of 2D and 3D
inversion codes. It also makes the interpretation of observations of lines formed in the chromosphere and transition region a slow and
computationally costly process, which limits the inference of the physical properties on rather small fields of view. Having access to a
fast way of computing the deviation from the LTE regime through the departure coefficients could largely alleviate this problem.
We propose to build and train a graph network that quickly predicts the atomic level populations without solving the non-LTE
problem.
We find an optimal architecture for the graph network for predicting the departure coefficients of the levels of an atom from
the physical conditions of a model atmosphere. A suitable dataset with a representative sample of potential model atmospheres is used
for training. This dataset has been computed using existing non-LTE synthesis codes.
The graph network has been integrated into existing synthesis and inversion codes for the particular case of \caii. We
demonstrate orders of magnitude gain in computing speed. We analyze the generalization capabilities of the graph network and
demonstrate that it produces good predicted departure coefficients for unseen models. We implement this approach in \hazel\ and
show how the inversions nicely compare with those obtained with standard non-LTE inversion codes. Our approximate method opens
up the possibility of extracting physical information from the chromosphere on large fields-of-view with time evolution.

\end{abstract}


\section{Introduction} \label{sec:intro}

The physical parameters of astrophysical objects are 
routinely inferred by interpreting 
the observed (polarized) spectrum in a selection of spectral lines. 
To this end, a parametric physical model, which is assumed to
be a good representation of the object under investigation, is proposed. This model
defines the temperature, density, velocity and magnetic field at a predefined
grid. This grid is three-dimensional in the general case, although two-
and one-dimensional grids are often used as simplifications.
Since the spectral lines are a consequence of transitions between pairs of
energy levels in specific atoms and molecules, one also needs to specify a model
atom/molecule. In the general case, the emergent polarization in a spectral line
is obtained by simultaneously solving the 
vector radiative transfer equation (RTE) for the Stokes parameters and the statistical
equilibrium equations (SEE) for the atomic level populations 
\citep{trujillo_manso99}. This self-consistent solution of the RTE and SEE is known as 
the non-local thermodynamic equilibrium (non-LTE) problem.
Since it is very difficult to solve, it is customary to make the simplifying assumption 
of neglecting atomic level polarization and work in what is commonly known as the Zeeman regime
\citep{landi_landolfi04}.

The non-LTE problem is nonlinear and nonlocal by nature. The nonlinearity appears because
the propagation matrix (that describes the absorption and dispersion of radiation) and the
emission vector that appear on the RTE depend on the atomic level populations.
Likewise, the matrix elements of the SEE depend on the radiation field at each
transition, which is obtained from the solution of the RTE. The nonlocality
comes from the coupling between distant points in the grid induced by 
the RTE. Due to the nonlinear and nonlocal character of the non-LTE problem, iterative methods 
are required to find the self-consistent solution of the RTE and SEE equations. The practical
majority of all methods in use at the moment are based on variations of point 
iterations with specific accelerations \citep[e.g.,][]{hubeny14}, being 
the accelerated $\Lambda$-iteration (ALI) the most widespread among all. Although methods with faster
convergence rates exist \citep[see, e.g.,][]{trujillo_fabiani95}, ALI turns
out to be ideal for its application to 3D problems.

The inference of the physical properties from the observation of spectral lines
can be carried out by inverting the RTE and SEE equations. Although one can carry out this
procedure by modifying by hand the proposed atmospheric model, it is much more efficient to rely on
automatic methods. These methods, commonly known as `inversion codes',
iteratively modify a guess model until the difference
between the radiation emerging from such model and the observations is small.
The distance between the synthetic and the observed Stokes profiles is typically
measured using the $\chi^2$ metric. The majority of automatic methods use the Levenberg-Marquardt
algorithm (a combination of a second order Newton method and a first order 
gradient descent method) to minimize the metric. To this end, they
make use of response functions, i.e., the elements of the Jacobian matrix describing how
the emergent Stokes profiles change when a perturbation is applied to the physical quantities.
The response functions are difficult to obtain due to the nonlinear
and nonlocal character of the non-LTE problem, and arguably the most convenient
way of computing them is through 
numerical evaluation using finite differences. As a consequence, each step of a non-LTE inversion requires  
many solutions of the full non-LTE forward problem (one for the synthesis and an
additional set for the calculation of response functions using finite differences).
It is apparent that a non-LTE inversion requires a significant amount of computation, 
even if, as shown by \cite{2017A&A...601A.100M}, the non-LTE response
functions can also be obtained efficiently using a semi-analytical approach.

The pioneering work of \cite{socas_navarro98} was the first to consider
the development of an inversion code for profiles in non-LTE. The code NICOLE 
was the first inversion scheme that could retrieve the stratification of physical properties from 
the interpretation of strong chromospheric spectral lines, which are commonly affected by non-LTE effects.
Now, after more than two decades, we have two more codes at our disposal. 
\citet{2018A&A...617A..24M} developed SNAPI, a non-LTE inversion code based on the
semi-analytical approach for the computation of the response function described in \cite{2017A&A...601A.100M}. \cite{2019A&A...623A..74D} developed the 
STockholm inversion Code (STiC), that upgrades the capabilities of previous
non-LTE inversion codes by allowing the inversion of 
very strong lines that require partial redistribution (PRD) in angle and
frequency. This is done in STiC by using the non-LTE synthesis
code RH \citep{uitenbroek01} as a 
forward module. This code allows the user to invert strong lines like the Mg \textsc{ii} multiplet,
emerging from the upper chromosphere. 

In contrast with the large computational requirements of non-LTE synthesis/inversion codes, 
the codes are much simpler when LTE conditions apply. LTE is reached whenever the source function is given by
the Planck function or collisional processes are much more important than
radiative processes. 
This assumption works, for instance, in many photospheric spectral lines, which are commonly used to analyze 
the deepest region of the solar photosphere. In this approximation, 
the propagation matrix and emission 
vector at each grid point depend only on local properties and can be expressed
with closed analytical formulae \citep{1992SoPh..137....1S,landi_landolfi04}. This induces
that the response functions can also be expressed in closed form. For this reason,
inversion codes based on the LTE assumption were developed well
before non-LTE inversion codes. Examples are the inversion codes SIR \citep{sir92} and
SPINOR \citep{frutiger00}. In general, SIR can carry out tens of synthesis per 
second, including the evaluation of the response function. A typical non-LTE forward synthesis can
require of the order of a few seconds to minutes, depending on the 
complexity of the problem. Therefore, a non-LTE iteration of an inversion 
code requires a factor 100 or more computing power than in the case of LTE.

Given this difference in speed, one would ideally like to have a non-LTE inversion code that performs
as fast as an LTE code. This is motivated by the fact that current telescopes routinely 
produce spectro-polarimetric observations with fields of view (FOV) of the order of megapixels, 
in many cases with time evolution. The time consuming non-LTE inversion codes are routinely only applied
to small subfields of the FOV and, even for this case, large supercomputers are required. 
From a theoretical point of view, a fast non-LTE inversion code could be obtained if one could have access, 
through an ``instantaneous'' computation (also known as oracle),
to the so-called departure coefficients, the ratio between the
populations in non-LTE and LTE. This is the motivation behind the
recently developed DeSIRe code \citep[Departure Coefficient 
aided Stokes Inversion based on Response functions;][in press]{desire_inpress}.
DeSIRe uses RH to compute the departure coefficients solving the full non-LTE forward
problem when the temperature of the guess model has changed more than a certain
threshold. In the rest of iterations, the inversion is performed with SIR assuming
LTE with fixed departure coefficients.

In this contribution we propose a new type of approach, in the form of a trained graph network \citep[GN;][]{2018arXiv180601261B}. Other similar approaches
based on machine learning are also under investigation (see Chappell \& 
Pereira, submitted).
The network uses the depth stratification of the physical conditions on an arbitrary
1D grid and produces the departure coefficients for a selection of atoms as output.
The graph network is fast and differentiable, so it can be used in combination with
an LTE inversion code to carry out non-LTE inversions much faster.
This opens the possibility of quickly getting chromospheric information from
large FOVs. We describe the architecture in the following, illustrating
its capabilities in the forward and inverse problem. We open source the
results in a public repository\footnote{\texttt{https://github.com/andreuva/graphnet\_nlte}} so that they can be used by the community.

\section{Bypassing the non-LTE problem with graph networks}\label{sec:neural_non-LTE}

\subsection{Radiative transfer: a short summary}
As described in the introduction, the standard multilevel radiative transfer problem 
requires the joint solution of the RTE equation, which
describes the radiation field, and the SEE for the atomic/molecular level populations, 
which describe the excitation state \citep[see, e.g.,][]{mihalas78}. We focus in
this work on its solution in 1D plane-parallel atmospheres in the unpolarized 
case, since our main aim is to build fast inversion codes. Anyway, our approximation can be 
easily extended to the general 3D case including atomic level polarization \citep[e.g.,][]{2021ApJ...909..183J}. We will investigate this in the future.

The basic physical properties of the atmosphere, (temperature, density, velocity, 
magnetic field, \ldots) are assumed to be known at a set of $N_\mathrm{P}$ discrete
points along the atmosphere. The objective is to compute the populations of
the $N_\mathrm{L}$ energy levels that are consistent with the radiation field 
within the stellar atmosphere. This radiation field has contributions from possible background
sources and from the radiative transitions in the given atomic/molecular model.

The SEE are given by the following expressions, which are formally 
a linear system of equations in $n_i$, the populations of the energy levels:
\begin{equation}
\label{eq:SEE}
\sum_{j \neq i}{n_j P_{ji}} - n_i \sum_{j \neq i}{P_{ij}} = 0.
\end{equation}
In this equation, $P_{ij}$ is the transition rate in s$^{-1}$ between level $i$ and $j$,
any of which can be bound or free. Two 
different processes contribute to this transition rate. Collisions between 
the model atom/molecule
and external abundant species induce collisional transitions between levels $i$ 
and $j$ with a rate given by $C_{ij}$. Additionally, 
radiative processes, denoted as $R_{ij}$, produce transitions between energy 
levels following certain selection rules. The total transition 
rate is then the sum of both terms $P_{ij}=C_{ij}+R_{ij}$. 

The collisional rates $C_{ij}$ only depend on the local physical 
conditions. On the other
hand, the radiative rates $R_{ij}$ depend on the radiation field, which
is a nonlocal quantity. For bound-bound transitions, we have:
\begin{eqnarray}
\label{eq:radiat_rates}
R_{ij} &=& B_{ij} \bar J_{ij} \qquad \qquad \quad i < j \nonumber \\
R_{ji} &=& B_{ji} \bar J_{ji} + A_{ji}\qquad i < j,
\end{eqnarray}
where $B_{ij}$ and $B_{ji}$ are the Einstein coefficients for stimulated emission 
and absorption,
respectively, $A_{ji}$ is the Einstein coefficient for spontaneous emission and 
$\bar J_{ij}$ ($\bar J_{ji}$) is the mean
frequency-averaged intensity weighted by the absorption (emission) profile:
\begin{eqnarray}
\label{eq:mean_intensity}
\bar J_{ij}  &=& \frac{1}{4 \pi} \int \mathrm{d} \mathbf{\Omega} \int \mathrm{d} \nu \, \phi_{ij} (\nu,\mathbf{\Omega})
I_{\nu \mathbf{\Omega}} \nonumber \\
\bar J_{ji}  &=& \frac{1}{4 \pi} \int \mathrm{d} \mathbf{\Omega} \int \mathrm{d} \nu \, \psi_{ij} (\nu,\mathbf{\Omega})
I_{\nu \mathbf{\Omega}}.
\end{eqnarray}
The symbols $\phi_{ij}$ and $\psi_{ij}$ stand for the normalized line absorption and 
emission profiles, respectively. In the complete frequency redistribution (CRD)
approximation, $\bar J_{ij} = \bar J_{ji}$ is fulfilled because the line emission 
and absorption profiles are the same.  $I_{\nu \bf \Omega}$ is the specific 
intensity at frequency $\nu$ and direction 
$\bf \Omega \rm$. The explicit directional dependence
in the line profiles account for directional Doppler shifts due to macroscopic velocity
fields in the medium. The integration has to be carried out for the full $4 \pi$ solid angle 
subtended by the sphere and for the frequency range of the transition. Similar
expressions are found for bound-free transitions but with the mean intensity
averaged over the photoionization cross-section, $\alpha_{ij}(\nu)$ \citep[see, e.g.][]{rutten03}.

The specific intensity is governed by the RT equation:
\begin{equation}
\label{eq:RT_eq}
\frac{\mathrm{d}}{\mathrm{d}s} I_{\nu \bf \Omega} = \chi_{\nu \bf \Omega} \left( S_{\nu \bf \Omega} -
 I_{\nu \bf \Omega} \right).
\end{equation}
which describes the variation of the specific intensity at frequency $\nu$ along a ray of direction
$\bf \Omega$. $\chi_{\nu \bf \Omega}$ and $S_{\nu \bf \Omega}$ are the opacity and the source function,
respectively, and $s$ is the geometrical distance along the ray \citep[see, e.g.,][]{mihalas78}.
The opacity describes how the photons are absorbed when being transported in the atmosphere, 
while the source function describes the generation of new photons in the atmosphere. 
The opacity and the source function have three main contributions. The first
one is coming from continuum transitions of background species.
We assume that this opacity can 
be obtained locally from the physical conditions at each spatial
point in the grid. This contribution includes, for example, the free-free and bound-free absorption 
processes in hydrogen, the background molecular species which are treated in LTE, opacity due to metals, 
absorption due to interstellar dust, etc. The second contribution is from 
the atomic/molecular model bound-bound transitions, that
can be computed from the population of the upper and lower levels. Finally, the
last contribution is due to continuum transitions in the atomic/molecular model.

Following \cite{rybicki_hummer92} \citep[see also][]{uitenbroek01,2021ApJ...917...14O}, 
we write
the emissivity ($\epsilon$) and the opacity ($\chi$) for a general 
transition between an upper level $j$ and a lower level $i$ as:
\begin{eqnarray}
\label{eq:opacity}
\chi_{ij}(\nu, \bf \Omega) &=& n_j U_{ji} \nonumber \\
\epsilon_{ij}(\nu, \bf \Omega) &=& n_i V_{ji} - n_j V_{ji}
\end{eqnarray}
The positive term in the opacity accounts for the direct absorption from the lower to 
the upper level and the negative 
term accounts for the stimulated emission from the upper to the lower level. 
The total emissivity and opacity for a given frequency is obtained by adding
together all sources. The expression for the $U$ and $V$ terms are the following:
\begin{eqnarray}
U_{ji}&=&
\left\{
\begin{array}{ll}
\frac{h \nu}{4 \pi} A_{ji} \psi_{ij}(\nu, \bf \Omega), & \quad \mathrm{bound-bound} \\
n_e \Phi_{ij}(T) \left( \frac{2h \nu^3}{c^2} \right) \mathrm{e}^{-h \nu/kT} 
\alpha_{ij}(\nu), & \quad \mathrm{bound-free},
\end{array}
\right. \\
V_{ij} &=&
\left\{
\begin{array}{ll}
\frac{h \nu}{4 \pi} B_{ij} \phi_{ij}(\nu, \bf \Omega), & \quad \mathrm{bound-bound} \\
n_e \Phi_{ij}(T) \mathrm{e}^{-h \nu/kT} 
\alpha_{ij}(\nu), & \quad \mathrm{bound-free},
\end{array}
\right. \\
V_{ji} &=&
\left\{
\begin{array}{ll}
\frac{h \nu}{4 \pi} B_{ji} \psi_{ij}(\nu, \bf \Omega), & \quad \mathrm{bound-bound} \\
\alpha_{ij}(\nu), & \quad \mathrm{bound-free},
\end{array}
\right.
\end{eqnarray}
where $n_e \Phi_{ij}(T)$ is given by the Saha-Boltzmann equation \citep[e.g.,][]{mihalas78},
with $n_e$ the electron number density.

The source function can be obtained as the ratio
\begin{equation}
\label{eq:source_function}
S_{ij}{\nu, \bf \Omega} = \frac{\epsilon_{ij}(\nu, \bf \Omega)}{\chi_{ij}(\nu, \bf \Omega)}.
\end{equation}
For our purpose, the previous equations can be more conveniently rewritten in terms of
the departure coefficient which, for any arbitrary level $k$, is given by
\begin{equation}
    b_k = \frac{n_k}{n^\star_k},
    \label{eq:dep_coef}
\end{equation}
where $n^\star_k$ is the population of level $k$ in LTE. This population can be easily
obtained by applying the Saha-Boltzmann equation, which only depends on local
quantities. For instance, the opacity for a bound-bound transition can be rewritten as
\begin{equation}
\label{eq:opacity_dep}
\chi_{\nu \bf \Omega} = \frac{h \nu}{4 \pi} b_l n^\star_l \phi_{\nu \bf \Omega} \left( 1 - \frac{b_u}{b_l} 
\mathrm{e}^{-h \nu/kT} \right),
\end{equation}
and the equivalent line source function as
\begin{equation}
\label{eq:source_function_dep}
S_{\nu \bf \Omega} = \frac{2h\nu_0^3}{c^2} \frac{1}{\frac{b_l}{b_u} \mathrm{e}^{h \nu/kT} - 1}\frac{\psi_{\nu \bf \Omega}}{\phi_{\nu \bf \Omega}},
\end{equation}
where $\nu_0$ is the central wavelength of the transition.

\subsection{Graph networks}\label{sec:graphnet}
Our objective is to define and train a neural network that can quickly produce
the departure coefficients for all the levels of the atom of interest at all
points in the grid. The architecture of the neural network has to fulfill some properties to be
of practical use. First, it needs to be fast. This is trivially fulfilled
by almost any architecture when compared with the computing time needed to solve
the non-LTE problem. Second, it needs to be flexible, so that one can apply it 
to arbitrary grids and not to a predefined one. Third, it needs to take into
account that the non-LTE problem is nonlocal and nonlinear. Therefore, the possible
correlation between the information from far apart grid points needs to be 
properly considered. An architecture that fulfills all these properties is the 
recently developed graph networks of \cite{2018arXiv180601261B}.\\

The fundamental idea consists in viewing the atmospheric grid in the atmosphere (1D, 2D or 3D)
as the connected graph $G=(V, E)$, where $V$ is the set of grid points (also 
known as nodes or vertices)
and $E$
is the set of edges connecting the grid points. This allows one to apply 
graph networks to an arbitrary number of grid points, placed at arbitrary locations in the 
atmosphere (so it is easy to work with non-cartesian atmospheres in
multiple dimensions). Each node represents the set of 
relevant physical properties $\mathbf{p}_i \in V$ at each location
of the atmosphere. Each edge $\mathbf{e}_{ij} \in E$ connects the two nodes $i$ (sender) 
and $j$ (receiver), and describes relevant inter-node properties.

\begin{figure}
    \centering
    \includegraphics[width=0.8\columnwidth]{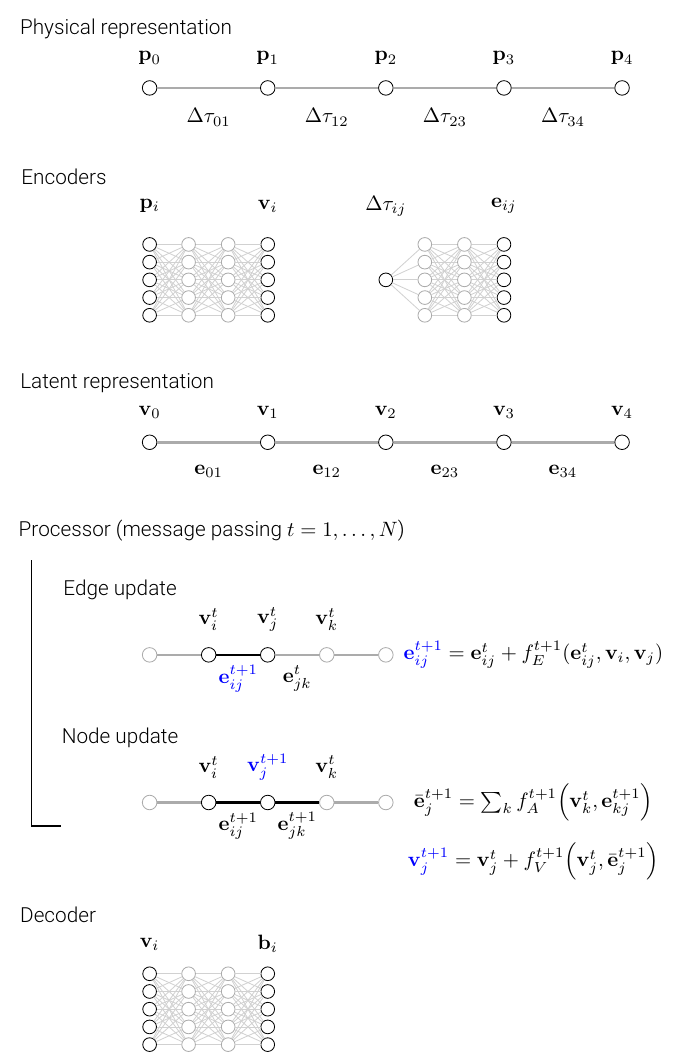}
    \caption{Schematic representation of the graph networks used to predict
    the departure coefficients. The quantities marked in blue in the
    processor phase are the ones being updated.}
    \label{fig:graphnet}
\end{figure}

A schematic representation of all the processing involved in a graph network,
as defined by \cite{2018arXiv180601261B}, is
shown in Fig. \ref{fig:graphnet}. The physical representation displays how a one-dimensional
model atmosphere is transformed into a connected graph. Nodes encode physical properties
$p_i$ and edges encode internode information. The second
step is to project the properties of the nodes and edges to a latent space of higher
dimension with the aid of two fully connected neural encoders. 
This projection largely increases the information encoding capabilities of the
graph network. For simplicity, we use the same dimensionality for both variables.
The resulting graph is shown in the latent representation, where nodes and
edges are labeled with their corresponding latent variables.

As described by \cite{2018arXiv180601261B}, the core of the graph network
occurs in the processor part, made of $N$ consecutive message passing processes. Each message
passing consists of updating the latent information contained in all edges and then in all nodes.
At iteration $t$, the edge update phase uses a fully connected neural network, $f_E^{t+1}$, to modify the
information encoded in the edge. The neural network accepts as input the concatenation of the 
edge and the nodes that are connected by the edge, and produces a modified output using an additional
residual connection \citep{residual_network16b}:
\begin{equation}
    \textbf{e}^{t+1}_{ij} = \textbf{e}^t_{ij} + f_E^{t+1}(\textbf{e}^t_{ij}, \textbf{v}^t_i, \textbf{v}^t_j).
\end{equation}
This very same neural network is applied to all edges (they can be done in
parallel) but it is different for each iteration of the message passing. 

The node update step modifies the node latent information by using a couple of fully connected networks, $f_A^{t+1}$ and $f_V^{t+1}$.
The first one is used to compute an effective edge information for each node. In general,
we use
\begin{equation}
    \bar \textbf{e}^{t+1}_{j} = \sum_k f_A^{t+1}\Big( \textbf{v}^t_k, \textbf{e}^{t+1}_{kj} \Big),
\end{equation}
where the summation is carried out over all edges $k$ for which the node $j$ is a receiver (in 
practice we use averaging instead of summation to reduce gradient explosion when backpropagating during training). 
A simpler form of this equation, which also gives good results \citep{DBLP:journals/corr/abs-2010-03409,DBLP:conf/icml/Sanchez-Gonzalez20},
is to simply average the value of the edges directly connecting other nodes to node $j$. In this case,
spatial invariance is introduced as an additional inductive bias, something that can 
be of interest in case one wants this symmetry to be conserved \citep[see][]{2018arXiv180601261B}.
Finally, the value of the node is updated by using:
\begin{equation}
    \textbf{v}^{t+1}_{j} = f_V^{t+1}\Big( \textbf{v}_j, \bar \textbf{e}^{t+1}_{j} \Big).
\end{equation}
Note that the encoding, decoding, node and edge update functions are
fully connected neural networks (FCNN), which take as inputs the physical quantities, 
node and/or edge
values and connect them to the desired output. Also note that the nodes at the extremes of
the graph have only
one edge. 

These FCNN use a combination of simple building blocks, containing
a series of learnable weights and biases, to approximate
a function or process through a carefully designed architecture 
\citep[see, e.g.,][]{chollet2017}. The basic building
block is the following:
\begin{equation}
    \textbf{a}^L = \sigma\left( \textbf{W}\textbf{a}^{L-1} + \textbf{b}^{L-1}\right),
    \label{eq:perceptron}
\end{equation}
where $\textbf{a}^{L-1}$ and $\textbf{a}^L$ are the input from
a previous layer and the output vector of the building block at 
layer $L$, respectively. The building block applies a matrix of 
weights, $\mathbf{W}$, to the input and adds a bias, $\textbf{b}^{L-1}$, and then
passes the result through a nonlinear activation function $\sigma$. The dimensions 
of $\textbf{a}^{L-1}$ and $\textbf{a}^L$ can be different depending on the size of the layers $L-1$ and $L$ respectively, so in general we have: $\textbf{a}^{L} \in \mathbb{R}^{n} $, $\textbf{a}^{L-1} \in \mathbb{R}^{m} $, $\textbf{b}^{L-1} \in \mathbb{R}^{n} $, $\textbf{W} \in \mathbb{R}^{n \times m} $. 

Message passing is fundamental to connect the information in very distant nodes in
the graph. Given that all updates are local (and can therefore be carried out in parallel), a simple
reasoning shows that $\sim N_\mathrm{p}$ iterations are needed to connect the information in 
all nodes in our simple graph topology. This number could be potentially reduced if a more
complex connectivity is used but its analysis is left for a future study.

A final decoder projects the values of the latent space in the nodes to the desired
output. In our case, we predict the departure coefficients for the $N_\mathrm{L}$ levels
of interest. This projection is again done in parallel for all nodes in the
graph.

\subsection{Specifics of GN for the non-LTE problem}

After a trial and error process, we decided to encode in the nodes
the temperature, $T$, in K, optical depth of the grid point (measured at 500 nm with 
$\tau_{500}=\int \chi_{500} \mathrm{d}s$), electron number density, $n_e$, in m$^{-3}$, microturbulent velocity, $v_\mathrm{mic}$, in m s$^{-1}$, and line of sight 
velocity, $v_\mathrm{LOS}$, also 
in m s$^{-1}$. We use the logarithm of all the
quantities, except for the velocities, to deal with the large variability.
The selection of these five variables has given very good
results as shown in this paper, but nothing fundamentally limits taking
into account more information 
in case of necessity. This would be the case of the presence of, for instance, 
strong magnetic fields, that would have
an important impact on the RTE and SEE (see Sec. \ref{sec:build_train}). 

In our case, for the edges, we only use the optical distance measured at 500 nm 
between the nodes as the encoded information. 
With this selection, $\textbf{e}_{ij}=-\textbf{e}_{ji}$, but
the formalism is general and can seamlessly deal with asymmetric edge properties.
For simplicity, we only consider edges that connect
consecutive nodes. The results shown in this paper demonstrate that this is a good
election, although we will explore more complex graphs structures in 
the future, where each node is
connected to the nearest $k$ neighbors. 

We point out that there is a certain degree of overlap on the information
encoded in the nodes and edges because the optical depth is encoded
in both. However, given the enormous variability in the optical 
depth axes for all the models that we consider for training and validation, we found 
that encoding its absolute value in the nodes gave better generalization
properties. We defer for the future potential simplifications of the node encoding to reduce the
overlap.

\section{Training}\label{sec:training}

\subsection{Building the training set}
\label{sec:build_train}
The graph network is trained with results from calculations performed with the 
Lightweaver framework \citep{2021ApJ...917...14O}. Given the
time consuming solution of the non-LTE forward problem, the ease of use of the
Lightweaver framework facilitates the deployment of the generation of the
training set in a parallel machine. Although the method is general, we focus 
here on results for \caii\, whose lines are widely used to infer the properties of the chromosphere. 
The \caii\ infrared triplet is routinely used to
extract quantitative information from the lower- to mid-chromosphere, given that plenty of
observations are available. In this work we make use of observations performed with the CRISP instrument \citep{2008ApJ...689L..69S} mounted 
on the Swedish 1-m Solar Telescope \citep[SST;][]{scharmerSST03,scharmer06}.

\begin{figure}
    \centering
    \includegraphics[width=\columnwidth]{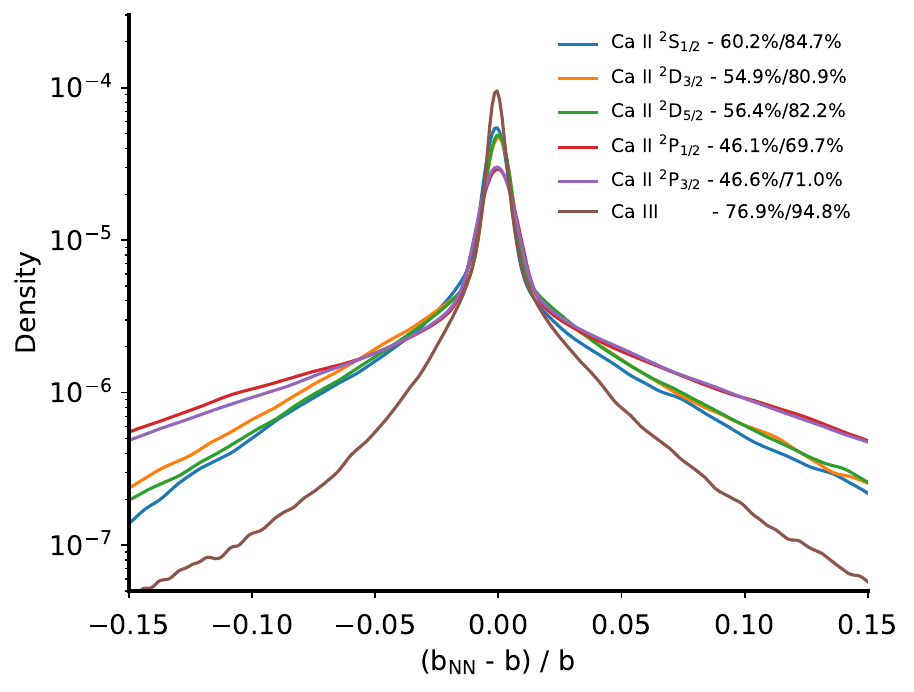}
    \caption{Gaussian kernel density estimation of the relative error between the correct
    departure coefficient and the one predicted by our neural approach for all 
    energy levels considered in the \caii\ model. The legend shows the 
    percentage of cases for which $b/b_\mathrm{NN}$ are in the intervals $[-0.01, 0.01]$
    and $[-0.05, 0.05]$, respectively.}
    \label{fig:residual}
\end{figure}

The model atom that we used is part of the Lightweaver distribution. 
We use a simple atomic model that contains the five bounded energy 
levels $4$s$^2$S$_{1/2}$, $4$p$^2$P$_{1/2}$, $4$p$^2$P$_{3/2}$, $3$d$^2$D$_{3/2}$ and 
$3$d$^2$D$_{5/2}$, responsible for 
the \caii\ infrared triplet and the H \& K lines in the blue part
of the spectrum. We also add the \caiii\
ground term to take into account all bound-free ionization transitions. The lines of the
infrared triplet are treated in CRD, while the H \& K lines are
treated in PRD. It is worth noticing that the CRD and PRD cases have different 
complexity levels. However, due to the large approximation capacity of the GN
we do not need to adapt the basic procedure, except perhaps increasing the complexity
of the GN by using more complex neural networks in cases in which the approximation error is
unacceptable large.


All calculations are made under the field-free simplifying
approximation for the sake of clarity, but it can be lifted if needed without fundamentally
affecting the procedure. 
This approximation assumes that the solution of the 
SEE is unaffected by the presence of Zeeman splitting in the energy levels \citep{1969SoPh...10..268R}.
If the magnetic field is very large, one should add its effect on the SEE. In this case,
apart from increasing the computation time for the generation of the database, one would probably
need to add the magnetic field strength (and potentially its orientation) as a new parameter in the nodes of the GN.

\begin{figure}
    \centering
    \includegraphics[width=0.8\columnwidth]{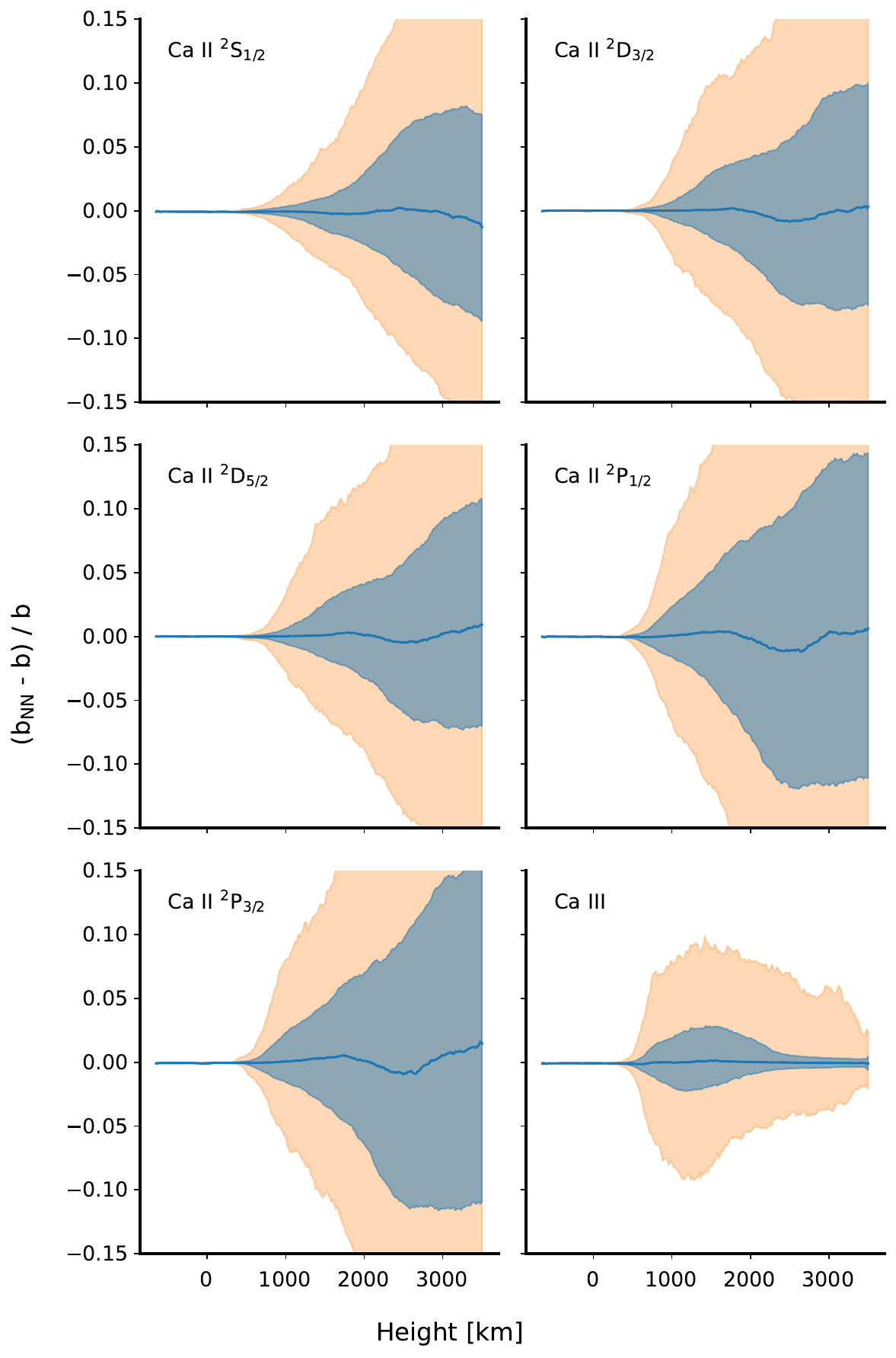}
    \caption{Geometrical height distribution of the relative error between the correct departure
    coefficient and the one predicted with the GN for all the \caii\ and \caiii\ atomic energy 
    levels. The median value is indicated with a solid
    blue line, while the shaded blue and orange regions indicate the 68\% and 95\% probability,
    respectively.}
    \label{fig:height_stats}
\end{figure}

With the sake of producing a GN that generalizes properly to new unseen models, we put some emphasis
on generating a database of atmospheric models with significant excursions on all
the physical properties. To this end, we use the semi-empirical 
FAL-A, FAL-C, FAL-F and FAL-XCO 
models \citep{fontenla_falc93} as baselines and add perturbations in the
temperature, microturbulent and line-of-sight velocities. We use signed Gaussian 
random perturbations with a standard deviation of 2500 K
for the temperature, a relative perturbation of 20\% for the microturbulent
velocity and 2.5 km s$^{-1}$ for the line-of-sight velocity. All these
perturbations are realized at nodes located at $\log \tau_{500}={-5,-4,-3,-2,-1,0}$.
Prior to adding them to the underlying model, these perturbations are 
adequately interpolated and smoothed to avoid 
very strong gradients that might compromise the convergence of the non-LTE problem. 
Since we have perturbed the temperature, the electron density of the model
is not anymore consistent. For this reason, the electron pressure is
computed for these models assuming hydrostatic equilibrium. To this set of 
randomized models, we add models extracted from the snapshot at $t=3850$ s of the
enhanced network simulation of \cite{2016A&A...585A...4C}, 
computed with the Bifrost code \citep{2011A&A...531A.154G}. The 
microturbulent velocity of these models is set to zero. A total
of 70 000 models have been produced in three datasets. 
One dataset contains 50 000 of those models and are actually used during 
training, while
the remaining 20 000 are used as validation and test sets to tune
the hyperparameters and check for over-fitting and generalization.


The full non-LTE problem for \caii\ is 
solved in each model atmosphere using Lightweaver using the MALI method with full preconditioning, similarly to RH.
The process is run in a distributed machine with 60 cores and takes $\sim 16$ hours.

\begin{figure}
    \centering
    \includegraphics[width=\columnwidth]{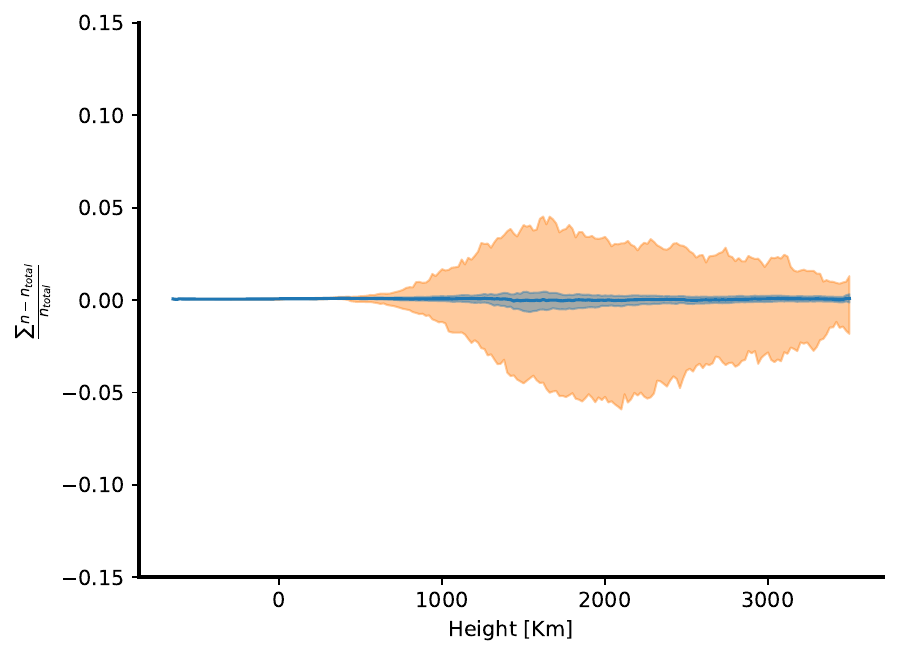}
    \caption{Height distribution of the relative error between the sum of the populations inferred with the computed departure coefficients and the true sum of populations for all the \caii\ and \caiii\ atomic energy 
    levels. The median value is indicated with a solid
    blue line, while the shaded blue and orange regions indicate the 68\% and 95\% probability,
    respectively.}
    \label{fig:populations}
\end{figure}

\subsection{Optimization}
There are a few hyperparameters that need to be defined a priori. The most important ones
are the size of the latent space used by the encoders (i.e., the dimension of vectors $v_i$ and $e_{ij}$),
the number of message passing iterations and the specific architecture of all fully connected
neural networks. We carried out a systematic parametric study and found some values that
seem optimal. The size of the latent space is set to 64. All neural networks, i.e., the
encoder, the decoder, $f_V^t$, $f_A^t$ and $f_E^t$, contain two hidden layers 
of size 64
with an ELU \citep{DBLP:journals/corr/ClevertUH15} activation function. 
Concerning the number of message passing steps, we find a significant
increase in precision when more message passing steps are used, albeit 
with a significant increase in 
memory consumption. For this reason, and because the improvement slows 
down significantly with more steps, we find a balance by setting
the number of message passing steps to 100.

The neural networks are coded in \texttt{PyTorch} \citep{pytorch19}, using the \texttt{PyTorch Geometric} package
\citep{Fey/Lenssen/2019} for neural networks in graphs. We use the Adam optimizer
\citep{Kingma2014} with a learning rate of $10^{-4}$, which is annealed 
at each epoch using a cosine rule. The training is done by optimizing the mean squared 
error (MSE) between the predicted and computed departure coefficients in logarithmic scale. 
We use a logarithm scale because the departure coefficients are 
spread over several orders of magnitude.
The number of total trainable parameters is 
$\sim 5.5$M.
We train for 400 epochs on an NVIDIA GeForce RTX 2080Ti, for a total of 52 h.
Due to GPU memory constrains, the batch size was limited to 32-64.

\section{Validation}\label{sec:validation}
\subsection{Departure coefficients}
Once trained, we checked the generalization properties of the GN, which
is fundamental in any approach based on
machine learning. To this end, 
we utilize
the snapshot at $t=5300$ s from the Bifrost simulation. This snapshot is extracted
24 min after the
one used for training. This time is a few times larger than the typical 
lifetime of a granule, so 
the new snapshot can be considered to be sufficiently uncorrelated with the one 
used for training.
We solve the non-LTE problem in all columns of this new snapshot and compare the results
with those of the neural prediction. After training, our baseline GN displays an average MSE loss 
computed over this test dataset of $10^{-4}$. Though this points out to a good generalization 
capabilities of the GN, we analyze the results in more detail.

Figure \ref{fig:residual} displays the probability distribution of the relative difference 
between the computed departure coefficient and the one predicted by the GN for 
all the atomic levels considered in the \caii\ model in a random 
subsample of 5 000 pixels. 
The legend also displays the percentage of pixels in the simulation with a relative
error below 1\% and 5\%, respectively. The distributions are very peaky (note that they are displayed in 
logarithmic scale), with a relatively Gaussian central core and extended
wings. These distributions figures clearly show the generalization capabilities
of the GN, giving predicted departure coefficients of significant precision in unseen data. 

\begin{figure*}
    \centering
    \includegraphics[width=\textwidth]{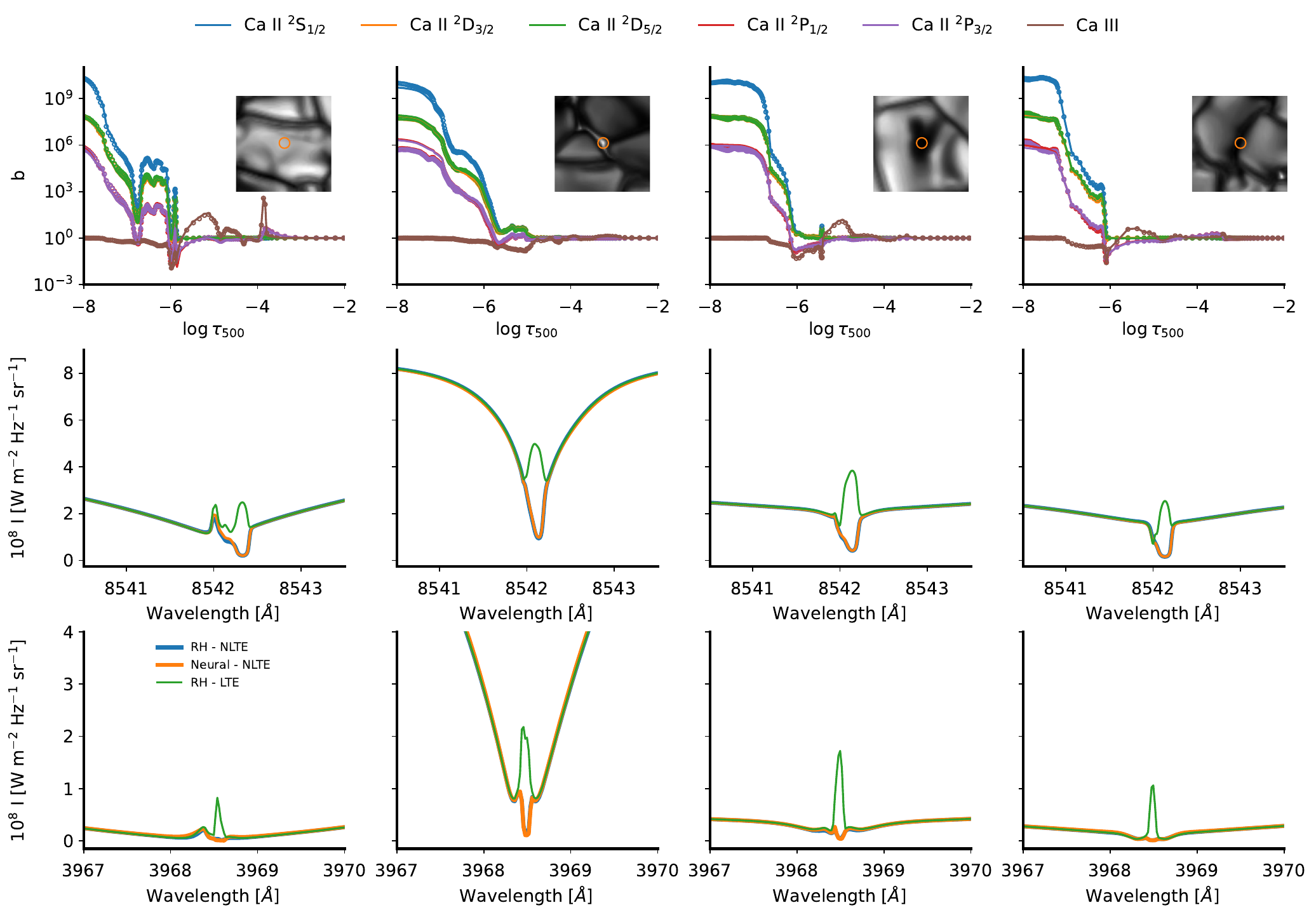}
    \caption{Upper panels: original (dotted) and approximated (solid) stratification
    of the departure coefficients with $\tau_{500}$ for
    all relevant levels of the \caii\ model atom on a few selected pixels from the Bifrost simulation
    snapshot shown in the right panel (we display the temperature map at $\tau_{500}=1$). 
    Middle panels: synthetic spectral line of the widely used \caii\ line at 854.2 nm using
    the original departure coefficients (blue) and the approximate ones (orange). For comparison, we also show the profile in LTE conditions in green. Lower panels: synthetic spectral line of the
    \caii\ K line at 396.8 nm, computed in PRD.}
    \label{fig:profiles}
\end{figure*}

This distribution is affected by the fact that the departure coefficients
in the deeper zones of the atmosphere are very close to unity, which our GN predicts
very well. Only a small fraction of all available heights really contribute
to the formation of the \caii\ lines. 
In order to better display this behavior, Fig. \ref{fig:height_stats}
shows the distribution of relative errors as a function of geometrical height in the 
3D snapshot. The solid blue line shows the median, while the shaded blue and orange
areas denote the $\pm 68$\% and $\pm 95$\% percentiles. It is clear that our prediction
for the deeper regions is always very good since all levels are in LTE. The quality of the
prediction decreases with height, reaching relative differences larger than 15\% in the
upper part of the snapshot. These heights correspond to the base of 
the corona or the upper part of the transition region, depending the selected pixel, but they are not important for the 
formation of any of the \caii\ lines. For the regions where the lines are formed, we
find relative errors that are safely below 5\% with 68\% probability.

It is important to point out that, in cases in which very precise departure coefficients are needed or to reduce
the probability of failure during the iterative solution, one can utilize the
GN approximate values as initial populations for solving the non-LTE problem. We have checked
that, using the ALI method, this amounts to a reduction in the number of iterations
needed for convergence of a factor between 2 and 4. Although relevant, the gain in computing time is not
specially important because ALI methods become slower when getting closer to the solution.

\begin{figure*}
    \centering
    \includegraphics[width=\textwidth]{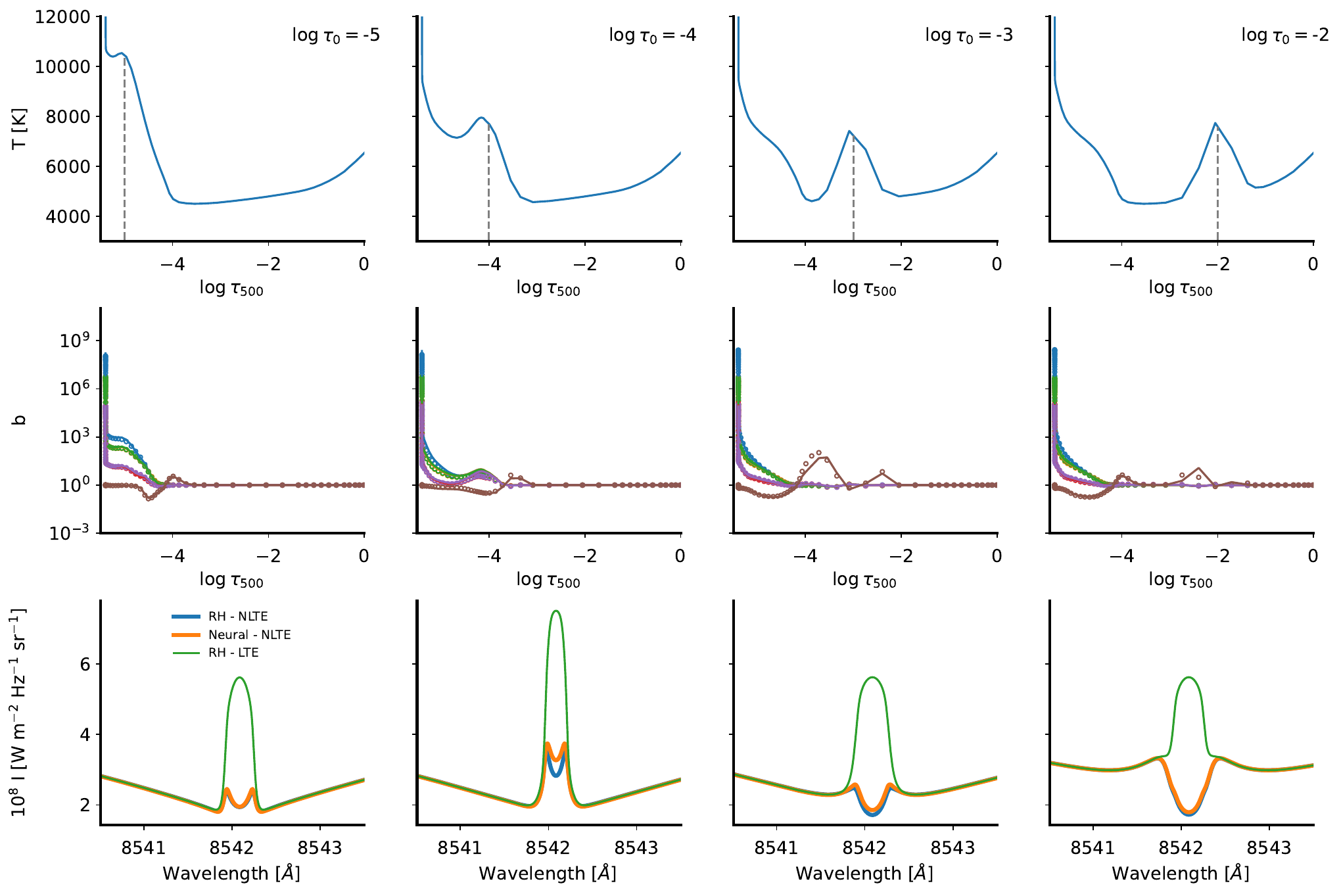}
    \caption{Generalization checks of a FAL-C model atmosphere to which a strong local
    increase in the temperature is added at different heights. The upper row shows the temperature
    stratification. The middle row shows the original (dotted) and approximated (solid) 
    departure coefficients. The lower row shows the synthetic lines using the correct
    departure coefficients (blue) and the approximated ones (orange), together with the
    profile in LTE (green).}
    \label{fig:bumps}
\end{figure*}

\subsection{Density conservation}
The atomic level populations computed with a non-LTE code fulfill, by construction, that
their sum is equal to the total number density of the specific atomic species. For the specific
case of our \caii-\caiii\ atomic model:
\begin{equation}
    \sum_i n_i = \sum_i b_i n^\star_i = n_\mathrm{Ca II} + n_\mathrm{Ca III},
\end{equation}
where we assume that the abundance of \cai\ is negligible. Since the GN
is used to predict $b_i$, we need to check at which level the previous
condition is fulfilled. Figure \ref{fig:populations} shows the relative error 
in the normalization of the populations that we incur when using the GN.
Most of the predictions give abundances that are consistent throughout the 
whole atmosphere. We find deviations clearly below 2\% with 68\% probability, and below
5\% with 98\% probability.


\subsection{Line profiles}
Although the GN produces a good approximation to the departure coefficient, the definitive
check requires the synthesis of the line profile. According to Eqs. (\ref{eq:opacity_dep}) 
and (\ref{eq:source_function_dep}), errors in the predicted departure coefficients
can be partially compensated. This is the case of the source function, where only the
ratio between departure coefficients appear, while a good prediction of the 
departure coefficient for the lower level is fundamental for a good estimation
of the opacity. In other words, one can still produce a good approximate line profile 
even if the approximate departure coefficients are not in excellent agreement with 
the correct one. To this end, we synthesized the line profiles for the \caii\ K (in PRD)
and 8542 \AA\ (in CRD) lines
at four representative locations in the 3D snapshot. We show the results
in Fig. \ref{fig:profiles}. The upper row
displays the calculated departure coefficients (dotted) for all levels of the \caii\ model, together with the
neural prediction. The precise pixel in the simulation can be seen in the inset.
Given the remarkable similarities on the departure coefficients, the level populations obtained from these
approximate departure coefficients produce synthetic spectral lines at $\mu=1$ (with $\mu$ the heliocentric
angle) in non-LTE that almost overlap with the 
full calculation, as can be seen in the middle (for the 8542 \AA\ line) and lower (for the K line)
rows of Fig. \ref{fig:profiles}. Note that the lines are asymmetric due to the
presence of strong velocity gradients with height in the selected pixels.
For comparison, we show in green the line obtained in LTE, which
is trivially obtained by setting $b_i=1$.

With our GN, the prediction of the departure coefficients takes around 
25 ms in a NVIDIA GeForce RTX 2080Ti, when done in serial. This includes all the input/output
operations required to load the input on the GPU memory and unload the results to the CPU memory.
Exploiting the parallelization properties of GPUs, a batch of roughly 50 such models
can be fit into typical GPU memory at once, producing similar computing times. On the other hand, the computation 
of the departure coefficients with Lightweaver can take 20 s in an Intel Xeon-6130 CPU @ 2.10 GHz. As a consequence, 
our approach produces a speedup factor of 10$^3$ with a negligible impact on precision. Arguably, larger speedups
can be obtained with smarter graph constructions that require less message passing steps.

As an additional test, we carry out the experiment
of computing the emergent line profile in the 8542 \AA\ line when a strong localized heating occurs
at different positions on the atmosphere. These models were not favored by the procedure followed
for the generation of the training set, so they are a good generalization test. Additionally, they
represent situations that surely happen in the solar atmosphere. The results are 
summarized in Fig. \ref{fig:bumps}. As shown in 
the upper panel of the figure, the proposed bump has Gaussian shape with an amplitude of 
3000 K and a full width at half maximum of 0.7 in 
$\log \tau_{500}$ units. It is added to the temperature stratification of the FAL-C model. The
central location of the heating is marked by the dashed grey line. The central panel shows 
the correct (dotted)
and approximate (solid) departure coefficients, with the same color code as that used in 
Fig. \ref{fig:profiles}. Finally,
the lower panel displays the emergent intensity profiles at $\mu=1$. The line profiles
are very similar to the correct ones, even though some differences exist in 
the central core. The differences are larger when the perturbations occur
on the line formation region. These differences should then be taken as representative 
of the uncertainty in the line profile introduced by the GN.

\begin{figure*}
    \centering
    \includegraphics[width=\linewidth]{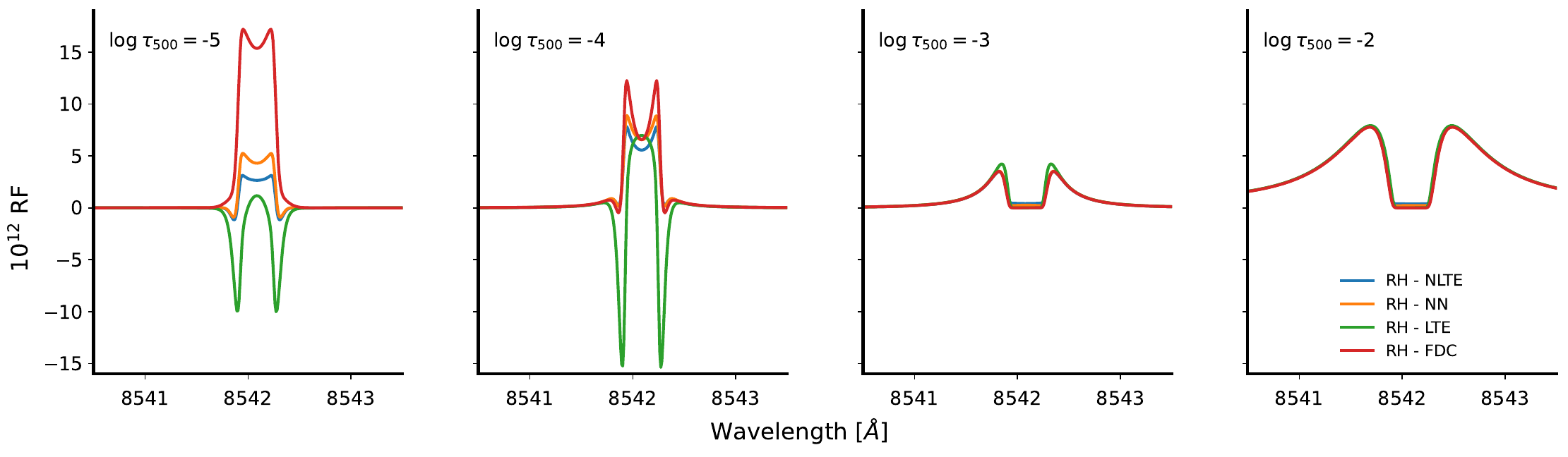}
    \caption{Response functions, in units of 10$^{-12}$, at four different optical depth heights
    for the FAL-C model atmosphere. The response functions are computed numerically using the full
    non-LTE calculation (blue), the approximate GN solution (orange), in LTE (green) and in the 
    fixed departure coefficient case (red).}
    \label{fig:response}
\end{figure*}

\begin{figure*}
    \includegraphics[width=\textwidth]{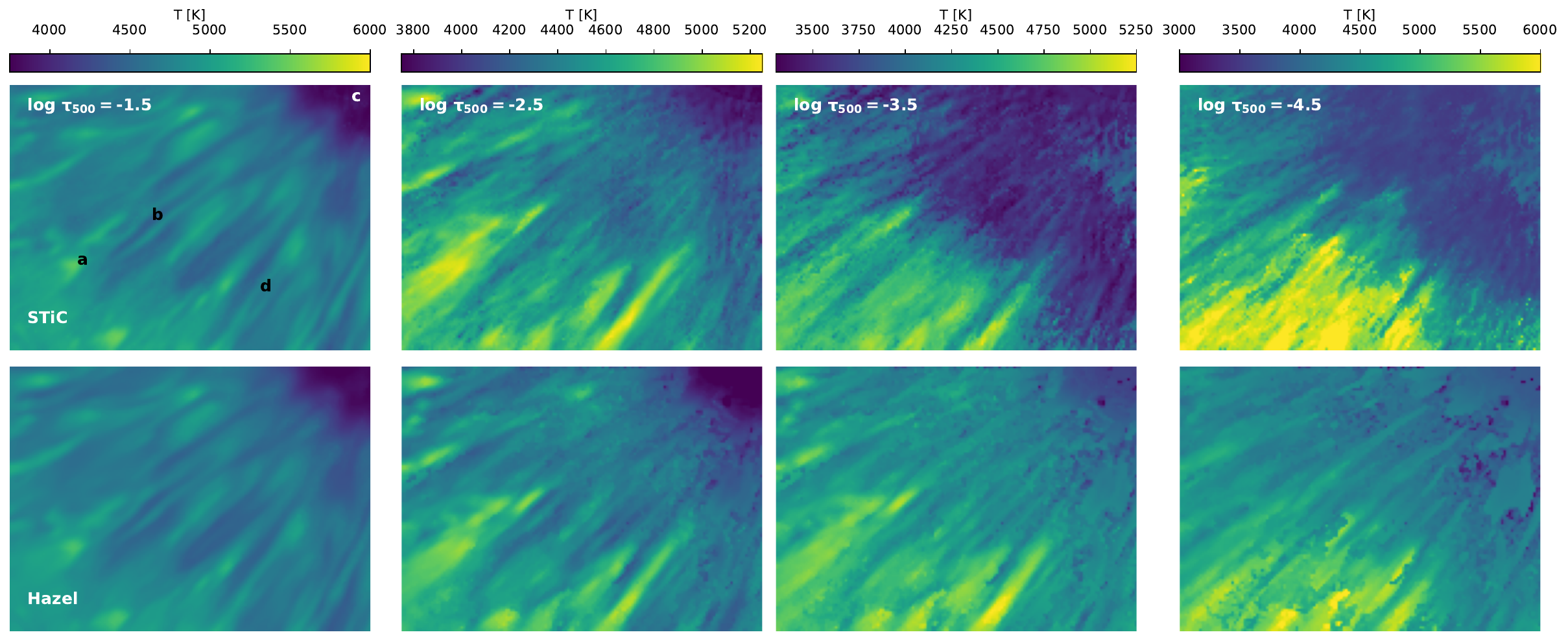}
    \caption{Temperature stratification at different optical depth surfaces
    obtained with STiC (upper panels) and \hazel\ (lower panel). The letters 
    refer to the position of the profiles displayed in Fig. \ref{fig:inversion_profiles}.\label{fig:inversion_T}}
\end{figure*}

\section{Inversions}
\subsection{Response functions}
As already mentioned, response functions are the derivative of the emergent line profile with respect to the 
physical conditions in the atmosphere. Having such precise Jacobian turns out to be crucial for the
development of a fast inversion code. 
We demonstrate with Fig. \ref{fig:response} that the GN approximation gives good
approximations to the response functions. The response functions are computed
in the FAL-C model on 6 nodes equispaced in $\log \tau_{500}$.
The temperature of each node is perturbed by a small amount ($\pm$25 K) and 
centered finite differences are used to compute the response functions. 
Four cases are considered in Fig. \ref{fig:response}. The first one (blue)
is the correct non-LTE response function computed with Lightweaver. The
second one (orange) uses the GN to compute the departure coefficients. The third case
(green) assumes that all departure coefficients are unity (LTE). The fourth case (red)
assumes that the departure coefficients are not affected by the perturbation and
are given by those of the unperturbed atmosphere. This is also known as the
fixed departure coefficients (FDC) case \citep{socas_navarro98}. The
FDC approximation clearly overestimates the response of the line to the temperature
at the higher layers, as shown by \cite{2017A&A...601A.100M}. The response function
is very similar to the correct one at lower layers for all approximations.
Our results clearly demonstrate that adding the dependence of the GN on the physical
conditions during the computation of the numerical derivatives leads to a very
good approximation of the response functions. This anticipates the very good
convergence results that we show in the following.

\subsection{Inversion results}
As discussed in the introduction, the GN serves as the perfect machinery to carry
out very fast non-LTE inversions. We implemented it into the \hazel\ inversion 
code \citep{asensio_trujillo_hazel08}
for the inversion of observations of the \caii\ 8542 \AA\ line. \hazel\ utilizes
the SIR machinery for the synthesis of lines in LTE. In order to synthesize
the line in non-LTE, the opacity and source function computed with SIR are modified 
according to Eqs. (\ref{eq:opacity_dep}) and
(\ref{eq:source_function_dep}) in the CRD approximation, with the departure coefficients computed
with the GN. Note that using the CRD approximation for the synthesis of the \caii\ 8542 \AA\ line 
is a good approximation because the GN returns the departure coefficients with the \caii\ H \& K
lines computed in PRD.

Since \hazel\ computes response functions using finite differences, and the GN
is differentiable by construction, the influence of the departure coefficients
are seamlessly included. This means that one is effectively working in the non-fixed departure
coefficient case. 
Although not strictly necessary, we checked that departure coefficients
do not need to be recomputed at every iteration. We empirically found good results by recomputing
them only when the maximum change in the temperature during the inversion procedure is larger than 50-100 K.

There is a significant computing time advantage of the non-LTE option of \hazel\ 
with respect to any other non-LTE inversion code. Since inversion codes are routinely run in parallel supercomputers,
the GN cannot be run in a dedicated GPU, in general. Therefore, we need to resort
to the less efficient CPU computation, which is roughly 10 times slower than a dedicated GPU.
We assume $\sim$250\,ms and $\sim$20\,s for the computation of the departure coefficients using a GN or the
ALI iterative scheme, respectively. Consequently, our method is approximately a factor $f=(t+20)/(t+0.2)$
faster, where $t$ is the time required for common tasks like computing hydrostatic equilibrium or solving the 
radiative transfer equation. Typical values are found to be $t<0.1$\,s, so that $f>67$.
In other words, one can carry out non-LTE inversions only slightly slower than LTE inversions.

\begin{figure*}
    \includegraphics[width=\textwidth]{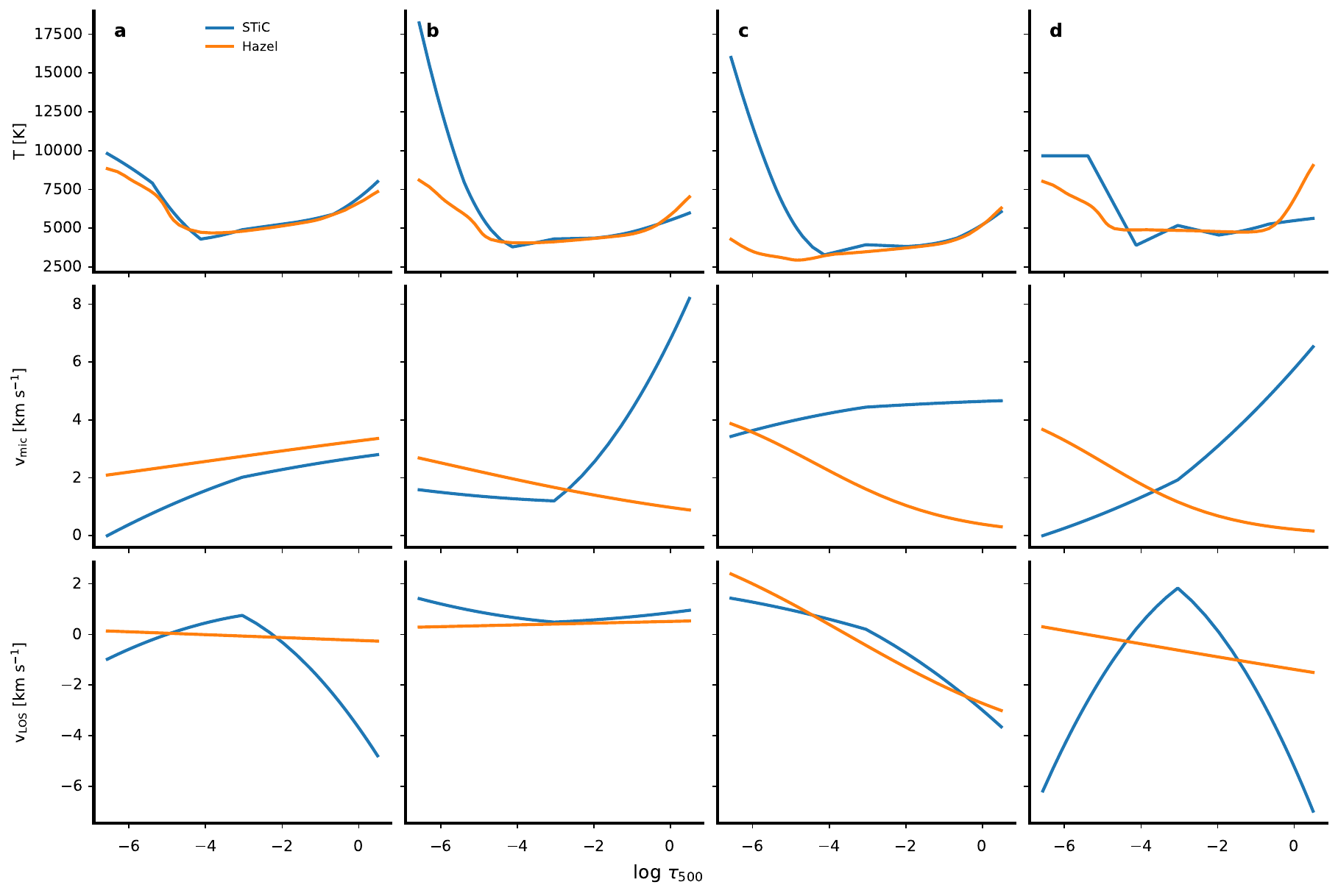}
    \caption{Depth stratification of the temperature (upper panel),
    microturbulent velocity (middle panel) and line-of-sight velocity (lower panel)
    for the pixels shown in Fig. \ref{fig:inversion_T} inferred with
    STiC (blue) and \hazel\ (orange).\label{fig:inversion_stratification}}
\end{figure*}

\begin{figure}
    \includegraphics[width=\columnwidth]{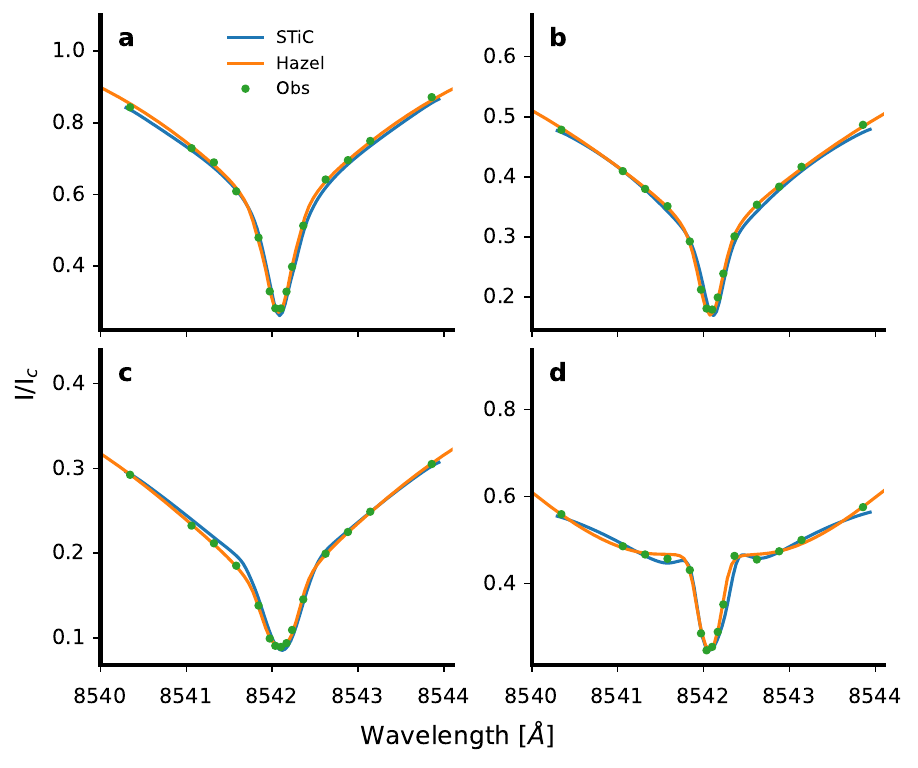}
    \caption{Observed Stokes $I$ profiles (green dots), together with
    the fits produced by STiC and \hazel. These profiles correspond to
    the pixel locations found in Fig. \ref{fig:inversion_T}.\label{fig:inversion_profiles}}
\end{figure}

The data that we invert consist of a scan from a time sequence of full-Stokes measurements 
of a penumbral region taken at high spatial resolution with CRISP in the \caii\ 8542 \AA\ line. 
The data were obtained on July, 27$\mathrm{^{th}}$ 2020, between 08:35:09 and 08:52:48 UT. At 
that time, the hosting sunspot was located at a heliocentric distance of 25.8 degrees. During 
the observation, the \caii\ 8542 \AA\ line was sampled at 15 wavelength positions in
the range $[-1.755,1.755]$~\AA, with steps of 65 m\AA \, close to the line center, and wider in 
the wings. Each scan was completed every 21 s. We acquired 12 accumulations per 
wavelength step and modulation state. The FOV of the observations 
is of about 55\arcsec $\times$ 55\arcsec, and the pixel size is 0\farcs057, which gives a total of 965 $\times$ 965 pixels. 
Due to the computational limitations of standard non-LTE inversions, we only analyze a small
subfield of 147$\times$107 pixels, equivalent to 8.3\arcsec $\times$ 6.1\arcsec. We 
carried out the data reduction using the SSTRED pipeline \citep{2015A&A...573A..40D,2021A&A...653A..68L}, 
and images were restored using the Multi-Object Multi-Frame Blind Deconvolution 
technique (MOMFBD; \citealt{2002SPIE.4792..146L}; \citealt{2005SoPh..228..191V}). 
The inversions are carried out both with STiC and \hazel, taking into account
the transmission profile of the CRISP instrument.

Given the differences between \hazel\ and STiC in terms of all the microphysics
and the inversion procedure, we do not expect both inversions to give exactly the 
same results. Since our interest here is to investigate how the GN works in non-LTE 
inversions, we put the emphasis on Stokes $I$ and the temperature. The STiC inversions
are done in three cycles. The first two cycles consider one every 10th pixel. We characterize
the configuration of each cycle with the vector $(n_T,n_v,n_\mathrm{vmic},n_B)$, which
gives the number of nodes in temperature, line-of-sight velocity, microturbulent
velocity and magnetic field components. The configuration for the first two cycles is 
(7,2,1,2) and (7,3,3,3). 
After the second cycle, we interpolate the results for all pixels and carry out a final cycle
with (7,3,3,3). In \hazel\, we simply invert every pixel from scratch using two cycles 
with (3,1,1,0) and (5,2,2,1) using FAL-C as a reference model. The average inversion time per pixel 
with \hazel\ in an 
AMD Epyc 7452 CPU is 9.8 s.

The temperature maps at four different optical depth surfaces are shown in 
Fig. \ref{fig:inversion_T}. The spatial appearance is remarkably similar between
both inversion codes, with clear chromospheric structures appearing
for layers above $\log \tau_{500} \leq -2$. The photospheric temperatures 
($\log \tau_{500} \geq -2.5$) are essentially indistinguishable, not only in their aspect
but also in the amplitude of the temperature. The results in higher layers ($\log \tau_{500} = -4.5$) 
are slightly different.
The information encoded in 
the profiles for such heights is reduced and the results depend on the 
specific assumptions of the model and the details of the inversion. 
This is not surprising since the CRISP data is undersampled, which
surely produces a loss of sensitivity around the central core, which
is sampling heights in the range $\log \tau_{500}=[-5, -4]$. For heights above
these, one should not put any emphasis on existing differences among codes because
the lines do not have any sensitivity \citep{2016MNRAS.459.3363Q}. Adding more
lines that help sample these regions is the only way to produce a more
constrained solution \citep{2018A&A...620A.124D}.

For a more in-depth analysis of the difference between the results of the
two inversion codes, we show in Fig. \ref{fig:inversion_stratification} the depth stratifications 
of the temperature, microturbulent velocity and 
line-of-sight velocity for the four pixels marked in Fig. \ref{fig:inversion_T}.
\hazel\ is based on the strategy proposed by \cite{sir92} of using a reference model as baseline.
This model is iteratively modified at the  nodes. It is obvious that the results of \hazel\ 
still display some memory from the FAL-C model used as reference. 
Meanwhile, STiC directly inverts the value of the physical properties at the nodes.
Anyway, the similarities between both results are obvious in the regions where there is 
sensitivity. The results are not expected to be exactly the same due to
the differences between the two synthesis codes, the different parameterization
of the solution and how the iterative inversion proceeds.
In the regions without sensitivity, both results might diverge but have negligible impact 
on the emergent profile. As discussed above, one should not focus on the differences in 
these regions of negligible sensitivity for the \caii\ 8542 \AA\ line.

Despite these differences in the stratification of the physical properties, the Stokes 
$I$ profiles are correctly reproduced with the two codes. Figure \ref{fig:inversion_profiles}
displays the Stokes $I$ profiles for the four selected points. The green dots show
the observed profile, with the orange and blue curves displaying the
\hazel\ and STiC synthetic profile in the inferred model, respectively.
Both fits display a similar quality.

\section{Conclusions and future work}\label{sec:conclusions}
We have shown that non-LTE problems can be solved very fast 
using a graph network for predicting the value of the departure
coefficients from the physical conditions. The forward evaluation of the
GN is orders of magnitude faster than the solution of the non-LTE problem. This
greatly accelerates the synthesis, opening up the possibility of quickly calculating
synthetic profiles of strong lines with non-LTE effects.
In cases in which the approximate departure coefficients are not precise enough, one
can use them as initialization for the full non-LTE problem. This leads
to a significant reduction on the computing time because fewer ALI
iterations (between half and one-fourth) are needed.

Arguably the most straightforward application of our GN is to accelerate non-LTE inversion codes.
As a proof of concept, we have implemented the neural approach in the \hazel\ inversion code, so that 
it can use the LTE machinery of SIR to carry out non-LTE inversions of \caii\ lines. We 
find that the computing time is reduced by a significant factor when compared
with classical approaches to non-LTE inversions. Given all the uncertainties of
a non-LTE inversion procedure (observational noise, lack of flexibility of the model, 
ambiguities, uncertainties in the atomic data,\ldots), our approximation turns out to be very 
competitive. The fact that we introduce a small error on the estimation of the
departure coefficients is only an extra component of this inherent uncertainty.

We demonstrated the fast inversions with data from the 
CRISP instrument showing a good comparison with the results obtained
with STiC.
We point out that the implementation in \hazel\ can only be used for lines for which the CRD
approximation is valid because it uses SIR internally. This is a good option for 
the lines of the infrared triplet of \caii. If lines 
need to be computed taking into account PRD effects, one needs to use codes like RH.
An acceleration of the inversion is still possible because the
departure coefficients can be computed with the GN and no MALI iteration is needed at all. We
plan to add this option to \hazel\ in the near future through the use of the Lightweaver
framework.

Although we have particularized all calculations to the prediction
of the departure coefficients in 1D plane-parallel cases, graph
networks are general. They can be seamlessly applied to higher dimensions
and to arbitrary topologies and for the prediction of other quantities. 
An obvious extension of this work is to compute 
departure coefficients or mean intensities in 3D snapshots, taking into 
account horizontal radiative
transfer effects. This would require the solution of the 3D non-LTE problem, 
something that can be done with codes like RH or MULTI3D \citep{2009ASPC..415...87L}.
The main difficulty resides on the computational effort to find a sufficiently large
training set.

We focus in this work on the computation of the departure coefficients for
the energy levels of the \caii\ model atom. However, the method is
general and can deal with any atomic system or even
combinations of them. Training such a GN could be of potential interest
to deal with recent multi-instrument inversions that use lines
from \caii\ with CRISP and CHROMIS on the SST, \mgii\ and Si \textsc{iv} with the
Interface Region Imaging Spectrograph \citep[IRIS;][]{2014SoPh..289.2733D} and 
Atacama Large Millimiter Array (ALMA) continua to map the chromosphere and
the transition region \citep[e.g.,][]{2019A&A...627A.101V,2020A&A...634A..56D}.

Despite the good results, some caveats are in order. First, 
our results are based on a pre-computed training set. If any of the
assumptions used during the computation of this set need to be changed, 
the trained GN is of no utility and a new training needs
to be carried out.
Second, although we have checked that the GN generalizes
correctly to some unseen models, the training set is obviously missing
models that describe energetic events, like those occurring in flares. 
We anticipate that model atmospheres
computed with the RADYN code \citep{1992ApJ...397L..59C,1997ApJ...481..500C}
can also be included in the training set to aid in the generalization properties
for this kind of energetic events. This would probably require the inclusion
of other physical quantities as new parameters of the nodes.
Additionally, the assumption of 
hydrostatic equilibrium can be lifted in these dynamic models.
Likewise, the Bifrost 3D snapshot corresponds
to a relatively quiet region, while active regions are not part of the
training set. Whenever a suitable MHD simulation of an active region including 
realistic chromospheres and transition regions is available, one can also add
them to the training set. Finally, we need to explore improvements to
the architecture and on the selection of hyperparameters to produce a GN with
lower generalization error in upper atmospheric layers.

\begin{acknowledgements}
We thank C. Quintero Noda for providing the 3D snapshots used in this
work. We thank J. de la Cruz Rodr\'{\i}guez, M. J. Mart\'{\i}nez
Gonz\'alez, T. del Pino Alem\'an, C. Quintero Noda, and J. {\v S}t{\v e}p\'an for
helpful suggestions to improve the paper. We acknowledge financial support from 
the Spanish Ministerio de Ciencia, Innovaci\'on y Universidades through project PGC2018-102108-B-I00 
and FEDER funds. This work was partly supported by the European 
Research Council (ERC) under the European Union's Horizon 2020 research and innovation programme 
(ERC Advanced Grant agreement No. 742265).
This paper uses data acquired at the Swedish 1-m Solar Telescope, operated on the island of La Palma by the Institute for 
Solar Physics of Stockholm University in the Spanish Observatorio del Roque de los Muchachos of the Instituto de 
Astrofísica de Canarias. The Institute for Solar Physics is supported by a grant for research infrastructures of national 
importance from the Swedish Research Council (registration number 2017-00625). The authors thankfully acknowledge the 
technical expertise and assistance provided by the Spanish Supercomputing Network (Red Española de Supercomputaci\'{o}n), as well 
as the computer resources used: the LaPalma Supercomputer, located at the Instituto de Astrofísica de Canarias.
This research has made use of NASA's Astrophysics Data System Bibliographic Services.
We acknowledge the community effort devoted to the development of the following 
open-source packages that were
used in this work: \texttt{numpy} \citep[\texttt{numpy.org},][]{numpy20}, 
\texttt{matplotlib} \citep[\texttt{matplotlib.org},][]{matplotlib}, \texttt{PyTorch} 
\citep[\texttt{pytorch.org},][]{pytorch19}.
\end{acknowledgements}


\bibliography{departure}{}
\bibliographystyle{aasjournal}



\end{document}